\documentstyle[amssymb,pra,aps,preprint]{revtex}

\input{tcilatex}

\begin{document}
\title{The superconducting gap of {\em in situ} $MgB_{2}$ thin films by microwave
surface impedance measurements}
\author{N. Hakim, C. Kusko, S. Sridhar.}
\address{Physics Department, Northeastern University, 360 Huntington Avenue,\\
Boston,MA 02115}
\author{A. Soukiassian, X. H. Zeng, X. X. Xi.}
\address{Department of Physics and Materials Research Institute, The Pennsylvania\\
State University, Universty Park, PA 16802}
\maketitle

\begin{abstract}
Precision measurements of the microwave surface resistance $R_{s}$ of in
situ $MgB_{2}$ films directly reveal an exponential behavior of $R_{s}$ at
low temperature indicating a fully-gapped order parameter. The entire
temperature dependence of $R_{s}$ is well described by a Mattis-Bardeen
formalism but with a small gap ratio of $\Delta (0)/kT_{c}=0.72$,
corresponding to $\Delta (0)=1.9meV$.

\newpage
\end{abstract}

Elucidation of the nature of the superconducting state in the novel binary
compound $MgB_{2}$ \cite{Akimitsu01} is of importance both for fundamental
understanding and for applications of this material in microwave
communications. Two important issues that are relevant are the nature of the
superconducting pairing mechanism and the symmetry of the order parameter.
There is mounting experimental evidence from different spectroscopies,
tunneling, specific heat and photoemission measurements that the
superconducting order parameter does not follow a BCS s-wave symmetry, and
presents a multicomponent gap\cite{Crabtree,Schmidt,Bouquet,SeneorChen}. In
the electromagnetic response, the characteristic signature of a gapped
superconductor is an exponential behavior of the microwave absorption
represented by the surface resistance $R_{s}$ at low temperatures.

We have measured the microwave properties of high quality $MgB_{2}$ films
prepared by Pulsed Laser Deposition (PLD) with an in situ annealing using a
new adaptation of a cavity perturbation technique for thin films in the
perpendicular geometry. Our results yield a direct measurement of the
superconducting gap and we find a value of $\Delta (0)/kT_{c}=0.72$. This
leads to a gap value of $\Delta (0)=1.92meV$ consistent with the smaller of
the two gaps typically observed in tunneling measurements. The entire
temperature dependence of the $R_{s}$ data is well described by a BCS s-wave
Mattis-Bardeen calculation with the same small gap ratio.

The films were grown by PLD deposition of $MgB_{2}/Mg$ mixture at 300 C
followed by an in situ annealing at 600 C\cite{XiArsen}. The structural
morphology of the films is nanocrystalline mixture of textured $MgO$ and $%
MgB_{2}$ with grain sizes around $50\stackrel{_{\text{ }o}}{A}$ (Fig.1).
Zero-resistance transition temperature of $34K$ was obtained in the best
samples using this technique, and the film used in this paper has a sharp
superconducting transition at $31K$ with width of 1K.

The measurements were carried out in a $Nb$ superconducting cavity resonant
at $10GHz$ using the ``hot finger'' technique which has been well
established to yield results on bulk and single crystals\cite{zhaisri0}.
Using a new analysis method, the microwave properties were measured in the
perpendicular geometry, i.e. $H_{\omega }\perp $ film plane. The film is
placed in the center of the cavity and the measurements are carried out in
the $TE_{011}$ mode. The sample temperature can be varied over a wide range
from $300K$ to $4.2K$ (or $2K$), while the $Nb$ cavity is maintained at $%
4.2K $ (or $2K$) ensuring a very high $Q$ and correspondingly low
background, and the surface impedance $Z_{s}=R_{s}+iX_{s}$ is obtained as
described below.

We have adapted the cavity perturbation method that has previously been
exhaustively analyzed and used for bulk samples \cite{zhaisri0}, to take
into account the enhancement of the external fields at the sharp edges of
the sample, we use a general theory by E. H. Brandt\cite{brandt1}
calculating the response to an applied perpendicular ac magnetic field $%
H_{\omega }$ of the thin film which is characterized by frequency dependent
complex resistivity $\tilde{\rho}_{\omega }$. In this formalism one is able
to calculate the sheet current $J_{\omega }(y,z)$ by solving a
one-dimensional integral equation. The solution can be approximated with
high precision by a finite sum and allows to calculate the magnetic moment
and the complex magnetic permeability $\widetilde{\mu }=\mu ^{^{\prime
}}-i\mu ^{^{\prime \prime }}$ of the sample. $\widetilde{\mu }%
=\sum_{n}c_{n}/(\Lambda _{n}+\tilde{w})$. The coefficients $c_{n}$ and the
poles $\Lambda _{n}$ were calculated by Brandt for both strip and disk
geometry \cite{brandt1}. The parameter $\tilde{w}$ is given by the
expression, $\tilde{w}=i\omega ad\mu _{0}\tilde{\sigma}_{\omega }(\omega
)/2\pi $ where $2d$ is the thickness of the sample and $a$ represents the
dimension of a square strip or the radius for a disk. The frequency
dependent complex conductivity is $\tilde{\sigma}_{\omega }=$ $\sigma
_{1}(\omega )-i$ $\sigma _{2}(\omega )=1/\widetilde{\rho }$

The methodology of the experiment is similar to the bulk samples microwave
measurements. At every temperature T, the cavity resonant frequency $%
f_{0}(T) $ and the width $\Delta f(T)$ were measured .The complex cavity
frequency $\tilde{f}\equiv f_{0}+i\Delta f/2$ is related to the sample
electromagnetic permeability $\tilde{\mu}_{\omega }$ by the following
equation: $\tilde{f}_{s}-\tilde{f}_{e}=g(1-\tilde{\mu}_{\omega })$, where $%
\tilde{f}_{e}$ and $\tilde{f}_{s}$ are complex frequencies of the empty
cavity and the loaded one respectively. From $\tilde{\mu}_{\omega }$ the
sample's complex conductivity $\tilde{\sigma}_{s\omega }$ can be determined
by inverting the complex frequency shift equation using a MATLAB routine.
The geometric factor $g$ was experimentally determined by assuming $g=\alpha
(f_{0s}(0K)-f_{0s}(T\gg T_{c}))$, where $f_{0s}(0K)$ was determined by
extrapolating $f_{0s}$ from the lowest measured temperature (usually $2K$ or 
$4.2K$) to $T=0K$. The coefficient $\alpha $ is determined such that the
calculated value of $\lambda (0)$ matches the experimental value. Only the
very low $T$ value of $\lambda $, viz.$\lambda (0)$ is sensitive to this
choice - overall the results obtained are extremely robust over a wide range
of $T$. From the measured complex conductivity $\tilde{\sigma}_{s\omega }$
the surface impedance $\tilde{Z}_{s}(\omega ,T)=R_{s}(T)+iX_{s}(T)=(\mu
_{0}i\omega /\tilde{\sigma}_{s\omega })^{1/2}$ was obtained. Note that this
is the bulk impedance of the thin film material, and not the actual thin
film impedance which can be calculated using suitable thickness-dependent
thin-film expressions.

The bulk superconducting microwave surface resistance $R_{s}(T)$ of the thin
film material is shown in Fig.2. The normal state surface resistance is
rather high $\sim 0.4\Omega $, consistent with a normal state resistivity of 
$230\mu \Omega -cm$. Although the surface resistance starts at relatively
high value above T$_{c}$, it starts to drop rapidly as temperature
decreases. At $T_{c}/2$, $R_{s}\sim 1m\Omega $ and below approximately $15K$
the drop becomes more rapid reaching values below $50\mu \Omega $ as the
temperature reaches $4K$. This last rapid decrease is characteristic of an
exponential behavior as will be seen shortly.

The bulk surface impedance of a BCS s-wave superconductor can be derived
using the relation between the impedance $Z_{s}=(-\mu _{0}i\omega /\tilde{%
\sigma}_{s})^{1/2}$ and the Mattis-Bardeen\cite{Mattis58} complex
conductivity $\tilde{\sigma}_{s}=\sigma _{1}-i\sigma _{2}$. The normalized
conductivity can be calculated using

\[
\sigma _{1}/\sigma _{n}=2/\hbar \omega \int_{\Delta }^{\infty
}g(E)[f(E)-f(E+\hbar \omega )]dE 
\]
, and

\[
\sigma _{2}/\sigma _{n}=1/\hbar \omega \int_{\Delta -\hbar \omega }^{\Delta
}\left[ 1-2f\left( E+\hbar \omega \right) \right] \left( E^{2}+\Delta
^{2}+\hbar \omega E\right) /\left[ \left( \Delta ^{2}-E^{2}\right) (\left(
E+\hbar \omega \right) ^{2}-\Delta ^{2})\right] ^{1/2}dE, 
\]
where $g(E)$ is the appropriate factor incorporating the density of states
and BCS coherence factors, and $f$ $(E)=[\exp (E/kT)+1]^{-1}$is the Fermi
function. The gap temperature dependence of the form $\Delta (T)=\Delta
(0)[1-(T/T_{c})^{2}]$ shown in the inset of Fig.2. We have used experimental
parameters $\omega =2\pi 10^{10}(\sec )^{-1}$ and $T_{c}=31K$. Using a small
gap ratio $\Delta (0)/kT_{c}=0.70\pm 0.02$, the calculated $R_{s}(T)$, is in
excellent agreement with the measured data over the entire temperature range
(Fig. 2).

The low temperature data shows an exponential temperature dependance
according to $R_{s}\varpropto \exp (-\Delta (0)/kT)$. To get a measure for
the gap we plotted $\ln R_{s}$ vs $T_{c}/T$ as shown in Fig.3. A straight
line behavior is clearly seen at low $T$, and is an unambiguous and direct
signature of the superconducting gap. A fit for $T\leq 15K$ gives a value of
the gap $\Delta (0)/kT_{c}$ $=0.72$, resulting in a gap value of $\Delta
(0)=1.92meV$. (We stress that Fig.3 represents the actual data, and that a
residual resistance was not subtracted , attesting to the high film
quality). The data represented in Fig.3 represent the clearest evidence of a
fully gapped superconductor from microwave measurements.

Although the gap ratio in the present work is smaller than the mean field
value of $1.76$, it is in good agreement with a number of reports on this
material \cite{Crabtree,Schmidt,Bouquet}. It appears fairly certain that $%
MgB_{2}$ is not a single gap s-wave superconductor - a tentative consensus
appears to favor a 2-gap superconductor characterized by $\Delta =\Delta
_{small}+\Delta _{l\arg e}$ or even a strongly anisotropic s-wave
superconductor with $\Delta =\Delta _{\min }(1+k\cos ^{2}\theta )$\cite
{Haas,SeneorChen}. Values for the smaller gap reported in the literature
vary between $\Delta _{small}(0)=1.9-2.8meV$ while for the larger gap $%
\Delta _{l\arg e}$the values vary between $6.2-9meV$. The microwave
measurements are sensitive to the smallest energy scale in the material and
hence we are clearly measuring the smallest gap $\Delta _{small}$ or $\Delta
_{\min }$. The good agreement with the calculations over the entire
temperature range indicates that the smaller gap ``turns on'' at $T_{c}$
along with a possibly larger gap which is not directly seen. This is
consistent with tunneling data which report that the small gap also appears
at $T_{c}$ as the temperature is lowered. The temperature dependence of the
gap indicated by our measurements is close to the mean-field temperature
dependence.

$MgB_{2}$ is of great interest as a superconductor with moderately high $%
T_{c}$ that might be suitable for electronic applications at $20K$ and
below. The $R_{s}$ values of the in situ film are quite low and competitive
with the best values of Y-Ba-Cu-O compounds at $4.2K$ and approach $Nb$
values. We have measured other ex situ films which had much higher values of 
$R_{s}$ likely due to their polycrystallinity and poor texture. Thus far the
present in situ films have yielded low $R_{s}$ and this may be attributed to
their nanocrystalline and highly smooth texture. The nanocrystalline
structure with $MgO$ contaminants may cause lower residual dc resistance
ratios of typically 1.4 for such films than in the good bulk samples and may
also result in the lower superconducting transition temperature than the
bulk since the grain size is comparable to the coherence length. At the same
time the nanocrystallinity appears to lead to some of the lowest microwave
losses observed \cite{Hakim,YoungLee,Andreone,Hein}.

The gap reported here is largest among the metallic and bimetallic
superconductors such as $Nb$ or $Nb_{3}Sn$ that are candidate materials for
microwave applications. If the normal state values are reduced, as is
implied by very low resistivities $<1\mu \Omega -cm$ observed in bulk
materials, then combined with the corresponding exponential drop in $R_{s}$
due to the fully gapped nature, dramatically low microwave loss ($%
10^{-9}\Omega -10^{-6}\Omega $) may be feasible at elevated temperatures ($%
20K$ and below) , not achievable with other high $T_{c}$ superconductors.
The present results show great promise for applications of $MgB_{2}$ in
microwave electronics at $20K$ and lower temperatures.

This work was supported at Northeastern by the Office of Naval Research
under grant N000140010002, and the work at Penn. State was supported by the
Office of the Naval Research under grant N00014-00-1-0294.

%
%
\begin{figure}
\caption{SEM of in situ $MgB_2$ film.}
\label{rsvsT}
\end{figure}

%

%
%
\begin{figure}
\caption{$R_s $ vs $T $ for $MgB_2$ The solid line represents the BCS calculation with
a gap parameter $\Delta(0) /kT_c = 0.70$.}
\label{rsvsT}
\end{figure}

%

\begin{figure}
\caption{$ln R_s$ vs. $1/T$. The line represents exponential behavior with a slope characterized
by $\Delta(0) /kT_c = 0.72$.
 }
\label{lnrsvs1/T}
\end{figure}%

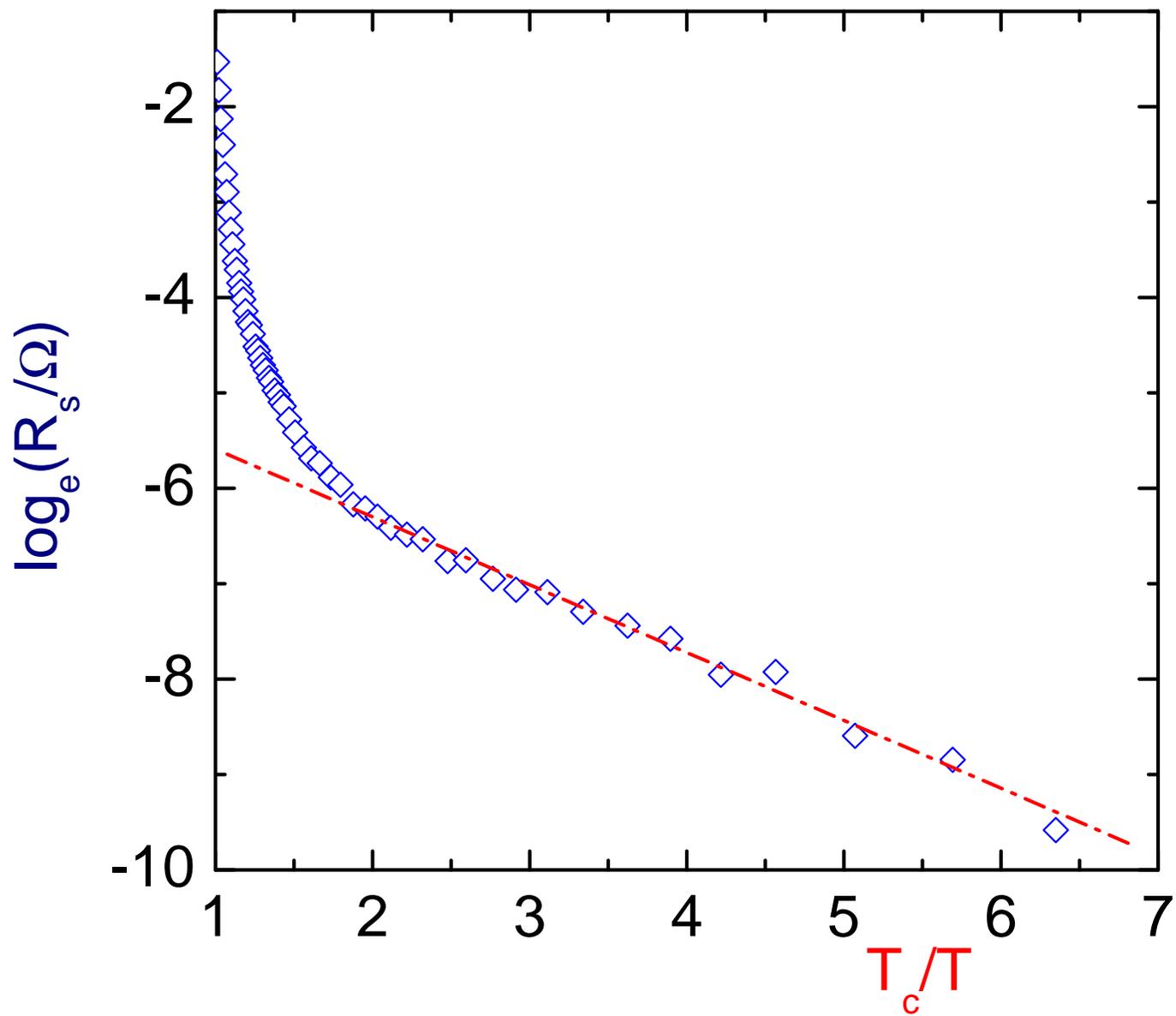
\end{document}